\documentclass[aip,reprint,amsmath,amssymb,citeautoscript,graphicx]{revtex4-2}

\usepackage{ulem}

\usepackage{dcolumn}
\usepackage{amssymb}
\usepackage{float}
\usepackage{graphicx}
\usepackage{booktabs}
\usepackage{multirow}
\usepackage{array}
\usepackage{makecell}
\usepackage{natbib}
\usepackage{fancyhdr}
\usepackage{comment}
\usepackage{bm}
\usepackage{soul}
\usepackage{amsmath}
\usepackage{xcolor}
\usepackage[colorlinks=true,bookmarks=false,citecolor=blue,linkcolor=blue,hyperfootnotes=true,urlcolor=blue]{hyperref}

\usepackage[utf8]{inputenc}
\usepackage{etoolbox}
\usepackage{mathptmx}
\usepackage[T1]{fontenc}

\begin{document}

\title{Heterostructure Engineering for Wurtzite LaN}

\author{A. J. E. Rowberg}
\affiliation{Quantum Simulations Group, Lawrence Livermore National Laboratory, Livermore, California 94550-9234, USA}
\affiliation{Materials Department, University of California, Santa Barbara, CA 93106-5050, USA}
\author{S. Mu}
\affiliation{Materials Department, University of California, Santa Barbara, CA 93106-5050, USA}
\affiliation{Department of Physics and Astronomy, University of South Carolina, Columbia, South Carolina, 29208, USA}
\author{C. G. Van de Walle}
\email[Author to whom to direct correspondence: ]{vandewalle@mrl.ucsb.edu}
\affiliation{Materials Department, University of California, Santa Barbara, CA 93106-5050, USA}

\begin{abstract}

Wurtzite LaN (wz-LaN) is a semiconducting nitride with favorable piezoelectric and ferroelectric properties, making it promising for applications in electronics.
We use first-principles density functional theory with a hybrid functional to investigate several features that are key for its use in heterostructures.
First, for the purposes of growing wz-LaN on a substrate or designing a heterostructure, we show that it can be lattice-matched with a number of cubic materials along their [111] axes.
We also evaluate the bound charge at such interfaces, taking into account both the polarization discontinuity and the piezoelectric polarization due to pseudomorphic strain.
Second, we investigate band alignments 
and assess the results for interfaces with zincblende-, rocksalt-, and perovskite-structure compounds, along with chemically similar wurtzite and rocksalt nitrides.
Our results provide guidance for the development of electronic devices based on wz-LaN.

\end{abstract}

\maketitle

\section{Introduction}
\label{sec:intro}

AlScN alloys are being heavily pursued for applications in electronics because of their attractive piezoelectric~\cite{akiyama2009enhancement,tasnadi2010origin,wang2021piezoelectric,hackett2023aluminum,alvarez2023thermal}, ferroelectric~\cite{fichtner2019AlScN,wolff2021atomic,guido2023thermal,chen2023scandium,Fichtner2024}, and dielectric~\cite{Casamento2022highK} properties.
These properties arise from AlScN having the wurtzite (wz) structure (at least up to a critical Sc concentration, which is on the order of 41\%~\cite{akiyama2009enhancement}).
ScN itself is most stable in the rocksalt (rs) structure (space group $Fm\bar{3}m$)~\cite{biswas2019development}.
Recently, other ferroelectric wz nitride alloys have been demonstrated, including 
AlBN~\cite{hayden2021ferroelectricity,savant2024ferroelectric}, GaScN~\cite{wang2021fully,uehara2021demonstration}, and AlYN~\cite{wang2023ferroelectric}.
Although they have not yet been explored, based on its position in the Periodic Table, one might expect similarly interesting properties to arise in alloys with lanthanum.

\begin{figure}
\includegraphics[scale=0.67]{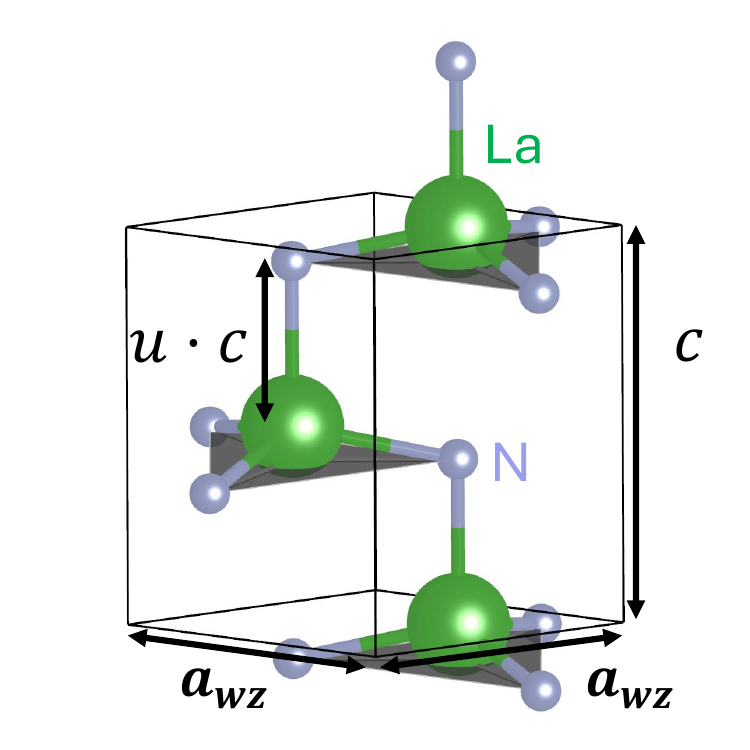}
\caption{Structure of wurtzite LaN, with lattice parameters $a_{wz}$ and $c$ labeled. The wurtzite $u$ parameter is also illustrated schematically. }
\label{fig:wz}
\end{figure}

Like ScN, YN and LaN are stable in the rocksalt phase (rs, space group $Fm\bar{3}m$).  
When attempting to stabilize them in a hexagonal unit cell, computational work has shown that ScN and YN do not adopt the wz structure but rather the layered, non-polar hexagonal structure (hex, space group $P6_3/mmc$)~\cite{farrer2002properties,cherchab2008structural}.
By contrast, for LaN, the polar wz structure (wz-LaN, space group $P6_3mc$) is locally and dynamically stable, and its total energy per formula unit (f.u.) is actually slightly lower than that of rs-LaN~\cite{ghezali2008structural,winiarski2019electronic,chen2021lan,rowberg2021structural}.
Experimentally, Krause $et$ $al.$ grew films of wz-LaN on a silicon substrate via reactive sputter deposition, providing evidence for its stability~\cite{Krause:nb5222}.
This wz structure is illustrated schematically in Fig~\ref{fig:wz}.
It is characterized by the $u$ parameter, defined as the ratio of the distance between a cation and the next-nearest anion plane to the $c$ lattice constant.
We previously calculated the $u$ parameter of wz-LaN to be 0.414~\cite{rowberg2021structural}.

In our recent computational study~\cite{rowberg2021structural}, we found that wz-LaN not only has significant spontaneous polarization (0.608~C/m$^2$), as expected for a wz structure, but also a high piezoelectric coefficient ($e_{33}=$1.78 C/m$^2$, compared to 1.57 C/m$^2$ for AlN).
These properties render the material attractive for devices such as high-electron-mobility transistors (HEMTs), in which the polarization discontinuity at a 
heterostructure leads to the formation of a two-dimensional electron gas (2DEG)~\cite{ambacher1999role,heikman2003polarization}.
We also demonstrated that the barrier for switching the polarization is low enough to enable ferroelectricity, potentially permitting applications in memory devices and ferroelectric field-effect transistors.
More recently, we also found that LaN has a high relative dielectric permittivity: $\varepsilon_{xx}(=\varepsilon_{yy})$=13.55 and $\varepsilon_{zz}$=16.21~\cite{rowberg2024point}.
It is enticing that enhanced piezoelectricity, permittivity, and potentially ferroelectricity can be obtained in a binary compound, as opposed to an alloy; for all these reasons, further investigations of wz-LaN and its relevant heterostructures are of high interest.

In this paper, we use first-principles calculations based on density functional theory (DFT)~\cite{hohenberg1964,kohn_self-consistent_1965} with a hybrid functional~\cite{heyd_hse} to evaluate properties of wz-LaN relating to heterostructures.
First, we survey materials that are reasonably closely lattice-matched to wz-LaN for use as substrates or in heterostructures, focusing on cubic materials where high-quality epitaxial growth is anticipated for $c$-axis growth along the cubic [111] direction.
We previously identified zincblende (zb, space group $F\bar{4}3m$) InP as a suitable candidate~\cite{rowberg2021structural}, but for many applications, a material with a larger band gap would be desirable.
Our survey of cubic materials with a lattice mismatch within $\pm$3.5\% to LaN leads us to identify a number of candidates.
Among those with wide band gaps, rs-NaBr and the cubic perovskite CsCaBr$_3$ stand out.
Among other candidates, the rs compounds GdSe, YAs, and GdAs have small lattice mismatch and may be suitable as substrates.
To assess the strain that would result from pseudomorphic growth, we calculate the elastic constants of wz-LaN.
These parameters allow us to include the piezoelectric contribution, in addition to the spontaneous polarization, in the calculation of polarization discontinuities (bound charge) at a heterostructure interface.

Second, we determine the band alignment of wz-LaN to other materials, employing the method of aligning the charge-state transition levels of interstitial hydrogen~\cite{van2003universal}.
We also obtain results for the alignment of wz-LaN to the other transition-metal nitrides, rs-ScN, rs-YN, and rs-LaN.

Throughout, we discuss our results in light of potential applications.
Among interfaces with zincblende materials, CdS is closely lattice-matched and exhibits a modest bound charge, which should make it feasible to modulate the resulting two-dimensional carrier gas in a transistor structure.
We also comment on alloys between wz-LaN and other III-nitrides, as they offer the potential of band-structure engineering along with tunability of the properties that render LaN promising.


\section{Computational Approach}
\label{sec:meth}

\subsection{First-Principles Calculations}
\label{ssec:firstprinciple}

Our first-principles calculations are based on DFT~\cite{hohenberg1964,kohn_self-consistent_1965} using the Heyd, Scuseria, and Ernzerhof (HSE)~\cite{heyd_hse,HSE06} screened hybrid functional, as implemented in the Vienna \textit{Ab initio} Simulation Package (VASP)~\cite{kresse_vasp}.
Use of a hybrid functional ensures a reliable description of the atomic and electronic structure.
The mixing parameter, which quantifies the fraction of non-local Hartree-Fock exchange, is set to its default value $\alpha=0.25$,
which has been shown to yield accurate values (within $\sim$0.1 eV) for the band gap of ScN~\cite{deng2015optical,mu2021firstprinciples}, a chemically similar material.
We use an energy cutoff of 500 eV for the plane-wave basis set, and the core electrons are described with projector-augmented-wave (PAW) potentials~\cite{Blochl_paw1,Kresse_paw2}, with the La $5s^2$ $5p^6$ $6s^2$ $5d^1$, N $2s^{2}$ $2p^{3}$, and O $2s^{2}$ $2p^{4}$ electrons treated as valence.
For other materials calculated explicitly in our study, we use PAW pseudopotentials that treat the Cs $5s^2$ $5p^6$ $6s^1$, Ca $3s^2$ $3p^6$ $3s^2$, Sc $3s^2$ $3p^6$ $4s^2$ $3d^1$, Y $4s^2$ $4p^6$ $5s^2$ $4d^1$, Se $4s^2$ $4p^4$, F $2s^2$ $2p^5$, and Br $4s^2$ $4p^5$, electrons as valence states.
An 8$\times$8$\times$6 $k$-point mesh is used to integrate over the wz-LaN unit cell.
Total energies are converged to within 10$^{-5}$ eV, and atomic relaxations are performed until forces are below 10 meV/\AA.
The elastic constants are calculated by extracting the stress-strain relationship from finite distortions of the lattice.

Hydrogen interstitials in the $+1$ and $-1$ charge states are calculated in supercells of wz-LaN, rs-ScN, rs-YN, and rs-LaN, as well as the cubic perovskite CsCaBr$_3$, for the purposes of computing band alignments.
We calculate their formation energies as:
\begin{equation}
	E^f({\rm H}_i^q) = E({\rm H}_i^q) - E_{\textrm{bulk}} - \frac{1}{2}E(\textrm{H}_2) + qE_F + \Delta_{\textrm{corr}} ,
\label{eq:form}
\end{equation}
where $E^f({\rm H}_i^q)$ is the formation energy of H$_i$ in charge state $q=\pm 1$; $E({\rm H}_i^q)$ is the total energy of a supercell containing H$_i^q$; $E_{\rm bulk}$ is the total energy of the defect-free supercell; $E$(H$_2$) is the total energy of a hydrogen molecule at 0 K; $E_F$ is the Fermi level; and $\Delta_{\textrm{corr}}$ is a finite-size correction term~\cite{freysoldt_2009}.

For wz-LaN, we use a 16-atom orthorhombic conventional unit cell to construct 2$\times$2$\times$2 supercells containing 128 atoms.
For rs structures, the 8-atom conventional cubic unit cell is used to construct 2$\times$2$\times$2 supercells containing 64 atoms.
For CsCaBr$_3$, the five-atom unit cell is used to construct 3$\times$3$\times$3 supercells containing 135 atoms.
A single special $k$-point is used for all supercell calculations.

\section{Results and Discussion}
\label{sec:res}

\subsection{Candidates for substrates and heterostructures}
\label{sec:het}

\subsubsection{Criteria for Heterostructures}
\label{sec:crit}

High-quality epitaxial growth of wz-LaN requires maintaining coherent chemical bonding at the interface, as well as reasonably close lattice matching.
This can be achieved using materials that have cubic symmetry, where the [0001] ($c$-axis) can be matched to a cubic [111] direction.
Lattice matching with other wz crystals may also be feasible, and in certain cases, these interfaces may also give rise to sizeable 2DEGs useful for HEMTs and similar applications~\cite{yoo2023microscopic}; however, wz-LaN has larger lattice parameters than most other wz materials, and therefore, compatible cubic structures are likely to be of more interest.

\begin{figure*}
\includegraphics[scale=0.34]{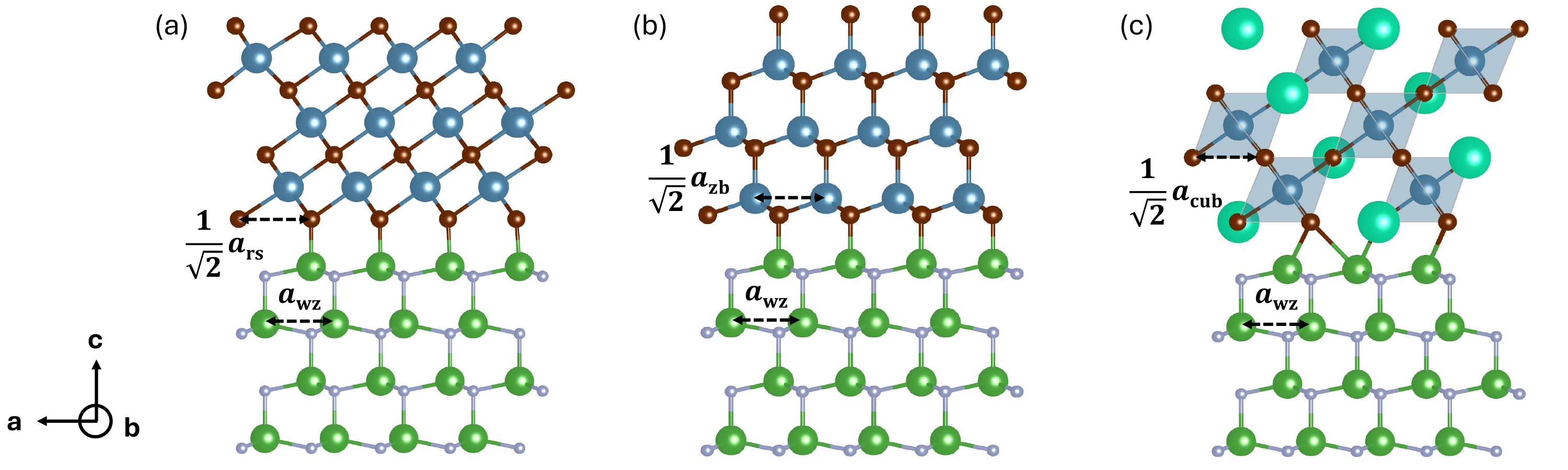}
\caption{Schematics of heterostructures between wurtzite LaN (bottom) and various cubic crystal structures: (a) zincblende, space group $F{\bar 4}3m$; (b) rocksalt, space group $Fm{\bar 3}m$; and (c) perovskite, space group $Pm{\bar 3}m$. }
\label{fig:hetero}
\end{figure*}

Interfaces between wz and cubic rs or zb materials are most easily envisioned by considering the stacking along the wz [0001] and cubic [111] directions, as shown in Fig.~\ref{fig:hetero}): in both cases, alternating layers of cations and anions are present, and the structure is determined by the relative position of these layers.
Along the [111] axis, rs systems [Fig.~\ref{fig:hetero}(a)] see cations coordinated with three anions each in the atomic layers immediately above and below them, while 
wz and zb [Fig.~\ref{fig:hetero}(b)] materials see cations coordinated with three anions in the layer below, and one anion in the layer above.
Note that wz [0001] and zb [111] are the ``cation-polar'' axes.
Both wz/rs~\cite{PERJERU2001490,Casamento2019} and wz/zb~\cite{spirkoska2009structural} heterostructures have been experimentally demonstrated.

Other cubic crystal structures can also be viewed in terms of stacking along [111]; e.g., the cubic ABX$_3$ perovskites (space group $Pm\bar{3}m$) [Fig.~\ref{fig:hetero}(c)] can be regarded as a stacking of AX$_3$ and B layers along the [111] axis.
We note that the B-site cations have the same sixfold anionic coordination as the cations in the rs structure.

As good lattice matching is important for high-quality heterostructures, either for purposes of growth or for use in an electronic device, we focus on the lattice constant of the cubic material and how it relates to the in-plane lattice parameter $a_{\rm wz}$ of the wz material; for that purpose, it is productive to define a ``cubic-equivalent'' lattice parameter:
\begin{equation}
    a_{\rm cub}=\sqrt{2}a_{\rm wz} \, .
\label{eq:cub-wz}
\end{equation}
For wz-LaN~\cite{rowberg2021structural}, $a_{\rm wz}=4.12$ {\AA} and thus $a_{\rm cub}=5.83$ {\AA}.
In similar fashion, equal surface areas can be identified for cubic and wz structures along the pertinent crystal directions.
These areas are $A=\frac{\sqrt{3}}{2}a_{\rm wz}^2$ for wz crystals along the [0001] direction, and $A=\frac{\sqrt{3}}{4}a_{\rm cub}^2$ for cubic materials along [111] \cite{adamski2020polarization}.
Comparing equal areas for distinct crystal structures is necessary to calculate polarization differences in heterostructures.

We previously identified zb-InP, with $a_{\rm cub}=5.87$ \AA, as being closely lattice-matched to wz-LaN~\cite{rowberg2021structural}; however, a material with a wider band gap may be more suitable for device applications.
For that reason, we conduct a search for suitable cubic semiconductors or insulators with wider band gaps using the Materials Project database~\cite{Jain2013,Jain2013commentary}.
Selected materials resulting from this search are listed in Table~\ref{tab:match}, along with their lattice constants and band gaps.
The structural data are based on a generalized gradient approximation (GGA) functional, as used by the Materials Project, which tends to overestimate lattice constants somewhat; for the purposes of our comparative search, these inaccuracies are unimportant. 
For band gaps, we report values that are obtained with a hybrid functional or experimental measurements to ensure a more accurate representation of the electronic structure.
We restrict our criteria for candidate materials to those with cubic symmetry having a volume within $\sim$10\% of that of the optimal lattice-matched material ($V=a_{\rm cub}^3=$198~{\AA}$^3$); i.e., we allow for a lattice mismatch of up to 3.5\%.
Such a mismatch is significantly better than that corresponding to silicon ($-$6.5\%), which has already been tried as a substrate for wz-LaN~\cite{Krause:nb5222}.
To narrow the design space further, we focus on materials with relatively abundant elemental constituents that are predicted to be stable.

We include a few additional chemically relevant materials in Table~\ref{tab:match} that do not meet our strain requirements, but are nonetheless of interest.
These include other wurtzite III-nitrides (AlN, GaN, and InN), as well as other IIIB-Nitrides (rs-ScN, rs-YN, and rs-LaN).
We will discuss these systems in more depth in our discussion of relative band alignments.

\begin{table*}
\setlength{\tabcolsep}{5pt}
\setlength{\extrarowheight}{4pt}
\caption{Cubic materials, grown along the [111] axis, with a close in-plane lattice match to wz-LaN. 
Other non-cubic materials of interest (specifically, other III-nitrides) are also included for reference. 
We quantify the expected lattice mismatch, which determines the strain in a wz-LaN layer grown on the material used as a substrate.
Lattice-constant data are taken from the Materials Project database~\cite{Jain2013,Jain2013commentary}, in which a GGA functional is used; each lattice constant listed reflects the cubic-equivalent lattice parameter.
Band gaps are taken from literature references, reflecting hybrid functional calculations where available; experimental data were used for GdP, CaSe, ZnSe, AlAs, GaAs, CdS, InP, Ge, AlN, GaN, and InN.
The band gaps for YSe, CsF, and CsCaBr$_3$ are calculated with the HSE hybrid functional in this work.
Where available, absolute positions of the valence-band maximum (VBM) and conduction-band minimum (CBM) relative to the vacuum level are provided; these values were calculated in this work (based on hydrogen alignment) if not otherwise indicated.}
\begin{tabular}{c|ccccc} \hline \hline
\thead{Material} & \thead{Cubic lattice parameter (\AA)} & \thead{Lattice mismatch, $\epsilon_1$} & \thead{Band gap (eV)} & \thead{VBM (eV)} & \thead{CBM (eV)} \\ \hline
 & \multicolumn{5}{c}{\bf Rocksalt, $Fm{\bar 3}m$} \\
ScN        & 4.51 & -29.7\% & 0.88 \cite{rowberg2021structural} & --5.71 & --4.83 \\
YN         & 4.90 & -19.4\% & 1.08 \cite{rowberg2021structural} & --6.00 & --4.92 \\
LaN        & 5.28 & -10.8\% & 0.76 \cite{rowberg2021structural} & --5.72 & --4.97 \\ 
CaS        & 5.70 & -2.6\%  & 3.57 \cite{chen2022design} & -- & -- \\
GdP        & 5.74 & -1.9\%  & 0.65 \cite{wachter1978electronic} & -- & --  \\
YSe        & 5.76 & -1.6\%  & 0 & -- & --  \\
AgBr       & 5.78 & -1.2\%  & 2.24 \cite{Peralta2006spin} & -- & -- \\
GdSe       & 5.82 & -0.5\%  & 0 \cite{Jain2013} & -- & --  \\
YAs        & 5.83 & -0.3\%  & 2.49\cite{kansara2017ab} & -- & --  \\
GdAs       & 5.89 & +0.7\%  & 0 \cite{khalid2020hybrid} & -- & --  \\
NaBr       & 5.92 & +1.2\%  & 7.18 \cite{miceli2018nonempirical} & -- & -- \\ 
CaSe       & 5.94 & +1.5\%  & 5 \cite{saum1959fundamental} & -- & -- \\
PbS        & 5.98 & +2.2\%  & 0.25 \cite{hummer2007structural} & -- & -- \\
CsF        & 6.01 & +2.7\%  & 7.17 & -- & -- \\ \hline

 & \multicolumn{5}{c}{\bf Zincblende, $F{\bar 4}3m$} \\
ZnSe       & 5.66 & -3.4\%  & 2.83 \cite{madelung2012semiconductors} & --5.96 \cite{van2003universal} & --3.13 \cite{van2003universal} \\
AlAs       & 5.68 & -3.0\%  & 2.23 \cite{madelung2012semiconductors} & --5.56 \cite{van2003universal} & --3.33 \cite{van2003universal} \\
GaAs       & 5.75 & -1.7\%  & 1.52 \cite{madelung2012semiconductors} & --4.97 \cite{van2003universal} & --3.45 \cite{van2003universal} \\
CdS        & 5.89 & +0.7\%  & 2.50 \cite{YU1992band} & --6.22 \cite{van2003universal} & --3.72 \cite{van2003universal} \\
InP        & 5.90 & +0.8\%  & 1.42 \cite{madelung2012semiconductors} & --5.16 \cite{van2003universal} & --3.74 \cite{van2003universal} \\ \hline

 & \multicolumn{5}{c}{\bf Diamond, $Fd{\bar 3}m$} \\
Ge         & 5.67 & -3.2\%  & 0.74 \cite{madelung2012semiconductors} & --4.27 \cite{van2003universal} & --3.53 \cite{van2003universal} \\ \hline

 & \multicolumn{5}{c}{\bf Perovskite, $Pm{\bar 3}m$} \\
CsCaBr$_3$ & 5.71 & -2.5\%  & 5.62 & --7.47 & --1.86 \\ \hline

 & \multicolumn{5}{c}{\bf Wurtzite, $P6_3mc$} \\
AlN        & 4.43 &  -24.4\%  & 6.10 \cite{YU1992band} & --7.43 \cite{van2003universal} & --1.33 \cite{van2003universal} \\
GaN        & 4.51 &  -22.9\%  & 3.50 \cite{YU1992band} & --6.73 \cite{van2003universal} & --3.23 \cite{van2003universal} \\
InN        & 5.02 &  -14.2\%  & 0.74 \cite{YU1992band} & --6.43 \cite{van2003universal} & --5.69 \cite{van2003universal} \\
LaN        & 5.85 &  0.0\%  & 2.29 \cite{rowberg2021structural} & --5.82 & --3.53 \\

\hline \hline
\end{tabular}
\label{tab:match}
\end{table*}

Again, schematics of the various interfaces that can be formed between wz-LaN and these cubic crystal structures are shown in Fig.~\ref{fig:hetero}.
We will analyze the properties of each of the wz-cubic interfaces separately, beginning with zb, proceeding with rs, and concluding with the cubic perovskite CsCaBr$_3$.

\subsubsection{Strain and Piezoelectric Polarization}
\label{sec:pz}

In addition to spontaneous polarization, piezoelectric polarization must also be considered when strain is present due to lattice mismatch in heterostructures.
Here we will assume the cubic material to be unstrained (i.e., acting as the substrate), while wz-LaN is pseudomorphically strained to match the in-plane cubic lattice parameter.
The piezoelectric polarization is given by~\cite{bernardini1997spontaneous,bernardini1999spontaneous,dreyer2016correct}:
\begin{equation}
    P_{\rm pz}=e_{31}\epsilon_1+e_{32}\epsilon_2+e_{33}\epsilon_3 \, ,
\label{eq:piezo}
\end{equation}
where $e_{31}=e_{32}=-0.12$ C/m$^2$ and $e_{33}=1.78$ C/m$^2$ are the relevant piezoelectric coefficients~\cite{rowberg2021structural}.
$\epsilon_1=\epsilon_2$ are the in-plane epitaxial strains (listed in Table~\ref{tab:match}), while $\epsilon_3$ is the strain along the $c$-axis, which can be expressed in terms of $\epsilon_1$
using the theory of elasticity:
\begin{equation}
    \epsilon_3=-2\epsilon_1\frac{c_{13}}{c_{33}} .
\label{eq:elastic}
\end{equation}

To evaluate these strains, we calculate the elastic constants of wz-LaN using HSE; these values are listed in Table~\ref{tab:elastic}.
Note that many of the values are equivalent by symmetry.
We observe that $c_{33}$ is significantly smaller than $c_{11}$, in contrast to other wz nitrides where these constants have very similar values~\cite{vurgaftman2003}.  
We attribute this contrast to the soft nature of the bonding along the $c$ axis in wz-LaN.
Indeed, the $u$ parameter of wz-LaN is 0.414~\cite{rowberg2021structural}, a value that deviates significantly from the $u$=0.375 value that would correspond to equal bond lengths for all bonds in the tetrahedral coordination environment. 
AlN, GaN, and InN all have $u$ parameters close to 0.375; the deviation in the case of wz-LaN indicates that the bond along the $c$ axis is significantly elongated, moving the structure closer to a centrosymmetric layered hexagonal structure.

\begin{table}
\setlength{\tabcolsep}{5pt}
\setlength{\extrarowheight}{4pt}
\caption{Calculated elastic constants for wz-LaN. Values not listed are zero by symmetry. }
\begin{tabular}{cr} \hline \hline
\thead{Elastic Constant} & \thead{Value (GPa)}  \\ \hline
$c_{11}=c_{22}$               & 140.8 \\
$c_{33}$                      &  89.2 \\
$c_{12}=c_{21}$               & 101.1 \\
$c_{13}=c_{23}=c_{31}=c_{32}$ &  84.5 \\
$c_{44}=c_{55}$               &  45.2 \\ 
$c_{66}$                      &  19.8 \\\hline \hline
\end{tabular}
\label{tab:elastic}
\end{table}

Adding the contributions from spontaneous polarization and piezoelectric polarization allows us to calculate the bound charge at various interfaces, $\sigma_{\rm b}$.
From Eq.~(7) in Ref~\onlinecite{dreyer2016correct}, the bound charge in our heterostructures based on pseudomorphically strained wz-LaN is calculated as:
\begin{equation}
    \sigma_{\rm b} = (P_{\rm sp}^{\rm sub}-P_{\rm sp}^{\rm wz-LaN}) - P_{\rm pz} = \Delta P_{\rm sp} - P_{\rm pz} .
\label{eq:bound}
\end{equation}
Here, $\Delta P_{\rm sp}$ is the difference in spontaneous polarization between wz-LaN ($P_{\rm sp}^{\rm wz-LaN}$) and the unstrained substrate material ($P_{\rm sp}^{\rm sub}$), while $P_{\rm pz}$ is the piezoelectric polarization of strained LaN, calculated using Eqs.~(\ref{eq:piezo}) and (\ref{eq:elastic}).
Our results are provided in Table~\ref{tab:bound}, with $\Delta P_{\rm sp}$ and $P_{\rm pz}$ listed separately to provide a sense for their relative importance in each heterostructure.
For consistency with Table~\ref{tab:match}, all values in Table~\ref{tab:bound} are calculated at the GGA level of theory; we verified that $P_{\rm sp}$ calculated at the GGA level is close to the HSE result, as exemplified by wz-LaN: 0.615~C/m$^2$ with GGA vs.\ 0.608~C/m$^2$ with HSE.

\begin{table}
\setlength{\tabcolsep}{5pt}
\setlength{\extrarowheight}{4pt}
\caption{Bound charges ($\sigma_{\rm b}$) resulting from polarization discontinuities ($\Delta P_{\rm sp}$) and piezoelectric polarization from pseudomorphic strain ($P_{\rm pz}$) in LaN at interfaces between wz-LaN and selected cubic materials. Each material's formal polarization ($P_{\rm sp}$) is also stated. Here, $P_{\rm sp}^{\rm wz-LaN}=0.615$ C/m$^2$, using GGA values.}
\begin{tabular}{c|cccc} \hline \hline
\thead{Material} & \thead{$P_{\rm sp}^m$ (C/m$^2$)} & \thead{$\Delta P_{\rm sp}$ (C/m$^2$)}  & \thead{$P_{\rm pz}$ (C/m$^2$)} & \thead{$\sigma_{\rm b}$ (C/m$^2$)} \\ \hline
 & \multicolumn{4}{c}{\bf Rocksalt, $Fm{\bar 3}m$} \\
CaS        & 1.139 &   0.524 &   0.092 &   0.431  \\
GdP        & 1.684 &   1.069 &   0.068 &   1.002  \\
AgBr       & 0.554 & --0.061 &   0.043 & --0.104  \\
YAs        & 1.633 &   1.018 &   0.012 &   1.006  \\
NaBr       & 0.528 & --0.087 & --0.044 & --0.044  \\
CaSe       & 1.049 &   0.434 & --0.056 &   0.489  \\
PbS        & 1.035 &   0.420 & --0.081 &   0.500  \\
CsF        & 0.512 & --0.103 & --0.099 & --0.004  \\ \hline

 & \multicolumn{4}{c}{\bf Zincblende, $F{\bar 4}3m$} \\
ZnSe       & 0.577 & --0.038 &   0.117 &  --0.155  \\
AlAs       & 0.860 &   0.245 &   0.105 &    0.140  \\
GaAs       & 0.839 &   0.224 &   0.061 &    0.163  \\
CdS        & 0.533 & --0.082 & --0.025 &  --0.057  \\
InP        & 0.797 &   0.182 & --0.031 &    0.213  \\ \hline

 & \multicolumn{4}{c}{\bf Diamond, $Fd{\bar 3}m$} \\
Ge         & 0.000 & --0.615 &   0.111 & --0.726  \\ \hline

 & \multicolumn{4}{c}{\bf Perovskite, $Pm{\bar 3}m$} \\
CsCaBr$_3$ & 0.000 & --0.615 &   0.072 & --0.687  \\

\hline \hline
\end{tabular}
\label{tab:bound}
\end{table}

\subsubsection{Heterostructures with Zincblende Compounds}
\label{sec:zb}

Viewed along the [111] direction, zb structures are characterized by an ABC stacking pattern, while wz structures along [0001] are characterized by AB stacking.
We have identified five zb materials with a close lattice match to wz-LaN: ZnSe, AlAs, GaAs, CdS, and InP.
We also include Ge here: it adopts the diamond cubic structure (space group $Fd{\bar 3}m$), in which the two atoms on the sublattices of zb are identical.
AlAs, GaAs, and InP are III-V semiconductors, which have a formal spontaneous polarization of 0.75 $e/A$ along the [111] axis, while ZnSe and CdS, as II-VI semiconductors, have a formal polarization of 0.5 $e/A$ along [111]~\cite{dreyer2016correct}.
Ge can be treated as a zb material with alternating layers of $+4$ and $-4$ charge, resulting in zero formal polarization.

In Table~\ref{tab:bound}, we quantify the polarization and bound charge associated with each of these zb systems in a heterostructure with wz-LaN.
The bound charge is largest for Ge, due to the large spontaneous polarization difference.
For ZnSe, the large pseudomorphic strain leads to a large contribution from piezoelectric polarization.
We note that Materials Project data reflected in Table~\ref{tab:match} significantly overestimate the lattice constant of GaAs as 5.75 \AA, whereas the experimental lattice constant is 5.65 \AA~\cite{LandoltBornstein2001}.
As a result, the actual piezoelectric polarization will be significantly larger than that reported in Table~\ref{tab:bound}; using the experimental constant would increase the contribution from piezoelectric polarization to 0.123 C/m$^2$, leading to an overall bound charge of 0.131 C/m$^2$. 

The smallest magnitude of the bound charge among zb candidates is found for CdS.
Small bound charges give rise to two-dimensional carrier gases that are easier to control and may therefore be useful for designing transistors.
CdS also has the closest lattice match among these zb materials ($+0.7$\% strain), making it an appealing candidate for heterostructure integration.

All six of these materials are semiconductors, but most have smaller band gaps than wz-LaN; the exceptions are CdS and ZnSe, which have slightly larger band gaps.

\subsubsection{Heterostructures with Rocksalt Compounds}
\label{sec:rs}

We apply a similar procedure to determine the bound charge present at the interface between wz-LaN and rs compounds, which while not necessarily well-tailored to heterostructure applications, may be useful as substrates (NaBr specifically has been used as a substrate~\cite{kiguchi1999interface}).
Along the [111] direction, rs materials have a nonzero formal polarization~\cite{adamski2020polarization,adamski2019giant}.
For I--VII rs compounds (AgBr, CsF, and NaBr), the formal polarization is 0.5 $e/A$;
for II--VI compounds (CaS, CaSe, and PbS), the value is 1.0 $e/A$; and for III--V compounds (GdP and YAs), it is 1.5 $e/A$~\cite{adamski2019giant,adamski2020polarization}.

A few of the rs compounds in Table~\ref{tab:match} (YSe, GdSe, and GdAs) are metals or semimetals. 
They could be used as substrates, particularly since GdSe and GdAs exhibit a small lattice mismatch with wz-LaN (within $\pm$1\%).
Semiconducting YAs is another candidate substrate, considering its exceptionally small misfit strain to wz-LaN (--0.3\% strain).

Formal polarization values and resulting bound charges for rs--LaN interfaces are also listed in Table~\ref{tab:bound}.
Bound charges for rs heterostructures are smallest for the I-VII compounds.
In particular, CsF gives rise to the smallest bound charge among all compounds we consider ($-0.004$ C/m$^2$), followed closely by NaBr.
These two materials also have large band gaps, preventing carriers from entering them from wz-LaN. 
For AgBr, the HSE-calculated gap is slightly smaller than that of wz-LaN (2.24 eV versus 2.29 eV), although experimental measurements show it to be slightly larger (2.69 eV~\cite{brown1973solid}). 

\subsubsection{Heterostructure with CsCaBr$_3$}
\label{sec:CsCaBr3}

Among the materials surveyed, CsCaBr$_3$ is the only example of a cubic perovskite that we found to be reasonably lattice-matched to wz-LaN.
CsCaBr$_3$ also stands out for its large band gap, which we calculated to be 5.62~eV using HSE.
The material has not been widely studied, although it has been explored as a scintillator, and single crystals are available~\cite{Grimm2006light,GRIPPA2013CsCaBr3}.
The cubic structure of CsCaBr$_3$ is stable from $-$34 $^\circ$C to its melting point of 831 $^\circ$C.

The formal polarization polarization of CsCaBr$_3$ along the [111] axis is zero (modulo the quantum of polarization).
Contributions from spontaneous and piezoelectric polarization to the bound charge are listed in Table~\ref{tab:bound}; the total bound charge at an interface between wz-LaN and CsCaBr$_3$ is --0.687 C/m$^2$.

\subsection{Band Alignment}
\label{sec:alignment}

The alignment of the conduction and valence bands is key to the applications of a heterostructure.
To determine the band alignments, we used the approach of Ref.~\onlinecite{van2003universal}, where it was found that the (+/$-$) charge-state transition level of interstitial hydrogen (H$_i$) can be used to align the band structures of semiconductors and insulators. 
Although other methods exist for calculating the band alignments (e.g., surface calculations to align bands to the vacuum level, or interface calculations between specific pairs of materials), the hydrogen alignment technique involves no interfaces or surfaces and reflects a ``natural band alignment,'' avoiding complications due to dipoles, lattice mismatch, and surface relaxation and reconstruction.

Calculating the alignment follows from calculating the formation energy of H$_i$ in the $+$ and $-$ charge states.
Since the hydrogen ($+$/$-$) level is located about 4.5 eV below the vacuum level~\cite{van2003universal}, it also offers a connection to other alignment methods.
We performed calculations for hydrogen interstitials in CsCaBr$_3$, wz-LaN, rs-ScN, rs-YN, and rs-LaN.
For other materials, we used results for band positions from Fig.~2 in Ref.~\onlinecite{van2003universal}.

\begin{figure*}
\includegraphics[scale=0.38]{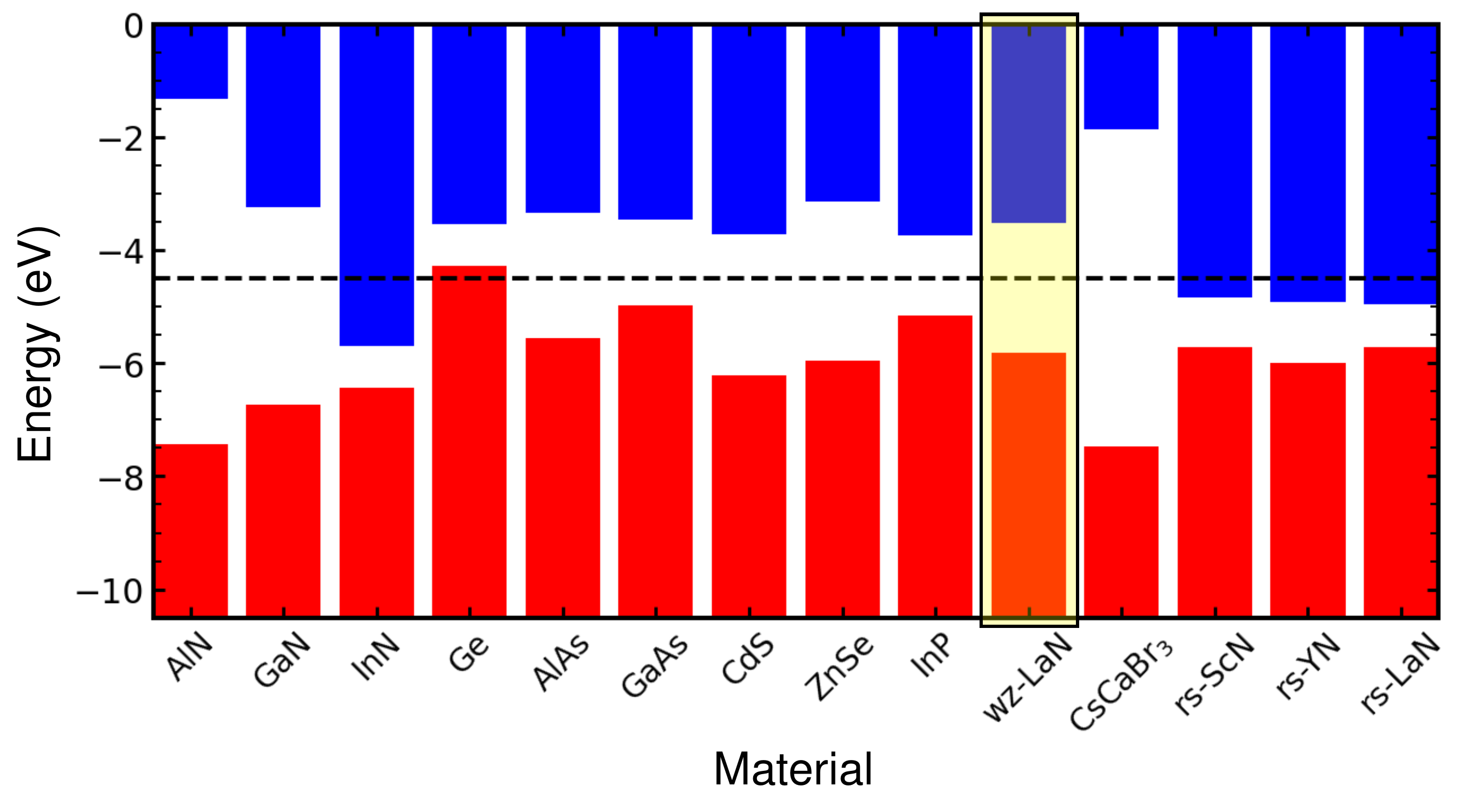}
\caption{Band alignments for materials relevant to this study, referenced to the vacuum level (zero of energy).
Valence bands are colored red and conduction bands blue.
The alignment for wz-LaN is highlighted for reference.
Alignments for wz-LaN, CsCaBr$_3$, rs-ScN, rs-YN, and rs-LaN 
were computed based on the charge-state transition level of interstitial hydrogen (horizontal dashed line) as described in the text;
other values for band positions were taken from Ref.~\onlinecite{van2003universal}.
VBM and CBM positions are listed in Table~\ref{tab:match}.  }
\label{fig:align}
\end{figure*}

The resulting alignments are shown in Fig.~\ref{fig:align}, and the VBM and CBM positions are listed in Table~\ref{tab:match}.
We first discuss alignment between wz-LaN and the conventional wz III-nitrides: AlN, GaN, and InN.\cite{Jain2013} 
These materials have significantly smaller $a_{\rm wz}$ lattice parameters than LaN (24\% smaller for AlN, 23\% smaller for GaN, and 14\% smaller for InN; see Table~\ref{tab:match}); the large lattice mismatch will prohibit formation of a coherent, pseudomorphic heterostructure with LaN.
Nonetheless, it is informative to consider the natural alignments, and the values could also be used for estimating alignments in alloys that will exhibit smaller size mismatches.
Both AlN and GaN have wider band gaps than wz-LaN and exhibit type-I (straddling gap) alignments, although the CBM of GaN is almost aligned with that of LaN.
InN has a much smaller band gap and a much lower position on the absolute energy scale, resulting in a type-II (staggered gap) alignment in which the CBM and VBM of InN lie below the corresponding bands of wz-LaN.

We also show alignments for the zb materials (along with Ge) that were identified as having a good structural match to wz-LaN.
The alignment between ZnSe and wz-LaN is type-I, 
with the band edges of wz-LaN lying within those of ZnSe.
For InP, the alignment is also type-I, but in contrast to ZnSe, the CBM and VBM both lie within the range of the band gap of wz-LaN, meaning that any carriers generated at the interface would reside in InP.
CdS, on the other hand, exhibits a type-II alignment, with both the CBM and VBM of CdS lying below the corresponding bands of wz-LaN.
AlAs and GaAs also exhibit type-II alignments, although their CBM and VBM levels lie above those of wz-LaN.
Finally, for Ge, the VBM is significantly above that of wz-LaN, but the calculated CBM positions are identical.
We note that these natural alignments do not include shifts in the band edges due to pseudomorphic strain.

On the right-hand side of Fig.~\ref{fig:align}, we show the band alignments for CsCaBr$_3$, rs-ScN, rs-YN, and rs-LaN, all of which arise from H$_i$ ($+$/$-$) calculations we perform here.
The alignment between wz-LaN and the much-larger-band-gap CsCaBr$_3$ is type-I.
For the rs nitrides, the hydrogen transition level lies above the CBM, and their CBMs are closely aligned, with a very slight monotonic decrease for rs-ScN$\rightarrow$rs-YN$\rightarrow$rs-LaN.
The position of the VBM does not change monotonically in the same fashion; instead, rs-YN has the lowest-lying VBM, slightly lower than that of rs-ScN, and the VBM of rs-LaN is about 0.3 eV higher.
This trend is largely---though not exclusively---due to the nature of the band gap changing from indirect for rs-ScN and rs-YN to direct for rs-LaN~\cite{rowberg2021structural}.
The alignment between rs-ScN and rs-LaN to wz-LaN is type-I, with wz-LaN having a slightly lower VBM and a significantly higher CBM; for rs-YN, however, the VBM is slightly below that of wz-LaN, leading to a type-II alignment. 

Band alignments can be tuned through alloying. 
For instance, in Al$_{1-x}$Sc$_x$N, the band gap can be changed dramatically over the range of $x$ for which Al$_{1-x}$Sc$_x$N maintains the wurtzite crystal structure \cite{deng2013bandgap}.
Similarly, we expect that wz-structure Al$_{1-x}$La$_x$N alloys can be formed over the entire composition range for $0 \leq x \leq 1$, with band gaps spanning a range of nearly 4 eV and offering impressive potential for band-alignment engineering.
Given that the piezoelectric coefficient $e_{33}$ of wz-LaN is higher than that of AlN, alloying La into AlN also offers the possibility of tuning piezoelectricity.
Furthermore, since we found the ferroelectric switching barrier to be modest in LaN~\cite{rowberg2021structural}, we anticipate that alloying La into AlN will facilitate ferroelectric switching.
Similar alloys of Ga$_{1-x}$La$_x$N and In$_{1-x}$La$_x$N should also be feasible, perhaps even more so due to the more similar cation sizes.
While their band gaps will be narrower than the gap of Al$_{1-x}$Sc$_x$N, they should offer similar benefits of piezoelectric tuning and ferroelectric switchability.

\section{Conclusions}
\label{sec:conclusions}

In summary, we have explored properties of wz-LaN that are pertinent for its applications in devices.
We considered various materials that might be used as substrates or in a heterostructure with wz-LaN.
We previously identified zb-InP as a well-lattice-matched material~\cite{rowberg2021structural}, albeit with a small band gap.
Here, we discussed a number of zb and rs semiconductors that also demonstrate small lattice mismatch with wz-LaN.
Among zb structures, CdS has a particularly close lattice match, and it also exhibits a modest bound charge; any resulting two-dimensional carrier gas should therefore lend itself to modulation in a transistor structure
Among rs structures, NaBr stands out because of its good lattice match, large band gap, and small bound charge.
In addition, we flagged the wide-band-gap cubic perovskite CsCaBr$_3$ because of its small lattice mismatch and the availability of single crystals.

We also presented results on the band alignment for wz-LaN with various relevant materials.
Type-I offsets, with wz-LaN band edges falling within a larger band gap, occur for heterostructures with ZnSe and CsCaBr$_3$.
Type-II offsets occur for AlAs and GaAs, where band edges lie above those of wz-LaN, as well as for CdS, where band edges lie below those of wz-LaN.
We also commented on the use of band-alignment engineering in case of alloying of wz-LaN with AlN, GaN, or InN.
We anticipate that our results will provide guidance for exploiting wz-LaN in applications that benefit from its piezoelectric, dielectric, and ferroelectric properties, as well as its large spontaneous polarization.

\begin{acknowledgments}

We acknowledge helpful discussions with Cyrus Dreyer.
A.J.E.R. was supported by the U.S. Department of Energy (DOE), Office of Science, Basic Energy Sciences (BES) under Award No.\ DE-SC0010689.
The work at the Lawrence Livermore National Laboratory was performed under the auspices of the U.S. Department of Energy (DOE) under Contract No. DE-AC52-07NA27344.
S.M. acknowledges the startup fund from the University of South Carolina and an Advanced Support for Innovative Research Excellence (ASPIRE) grant from the Office of the Vice President for Research at the University of South Carolina.
C.G.V.d.W. was supported by SUPREME, one of seven centres in JUMP 2.0, a Semiconductor Research Corporation program sponsored by the Defense Advanced Research Projects Agency. 
Computational resources were provided by the Extreme Science and Engineering Discovery Environment (XSEDE), which is supported by NSF Grant No. ACI-1548562, and by the DOD High Performance Computing Modernization Program at  the  AFRL  DSRC  and  ERDC  DSRC  under  Project  No. AFOSR46403464.
Use was also made of computational facilities purchased with funds from the National Science Foundation (CNS-1725797) and administered by the Center for Scientific Computing (CSC). The CSC is supported by the California NanoSystems Institute and the Materials Research Science and Engineering Center (MRSEC; NSF DMR 2308708) at UC Santa Barbara.

\end{acknowledgments}

\bibliography{wz-LaN}

\end{document}